\documentclass[11pt]{article}
\usepackage{moriondEprint,epsfig}
\usepackage{wrapfig,rotating}
\usepackage{enumerate}
\def\be{\begin{equation}}
\def\ee{\end{equation}}
\def\bea{\begin{eqnarray}}
\def\eea{\end{eqnarray}}

\newcommand {\pom} {I\!\! P}
\vspace*{-1.5cm}
\begin{document}
\vspace*{4cm}
\title{DIFFRACTION, SATURATION AND {\boldmath$pp$} CROSS SECTIONS AT THE LHC~\footnote{Presented at {\em XLVI~$^{th}$ Rencontr\`es de Moriond, QCD and High Energy Interactions, La Thuile,  Aosta Valley, Italy, March 10-27, 2011}.}}

\author{K. GOULIANOS}

\address{The Rockefeller University, 1230 York Avenue, New York, NY 10065-6399, USA}

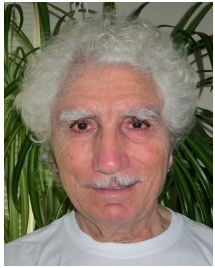
\maketitle\abstracts{Results from the large hadron collider (LHC) show that no available Monte Carlo simulation
 incorporates our pre-LHC knowledge of soft and hard diffraction in a 
way that could be reliably extrapolated to LHC energies. As a
 simulation is needed to establish triggers, perform underlying event
corrections and calculate acceptances, the lack of a robust simulation
affects all measurements at the LHC. Particularly affected are the measurements
of processes with large diffractive rapidity gaps, which constitute about one quarter of the 
inelastic cross section. In this paper, a previously described phenomenological model based on a saturation effect observed in single diffraction dissociation in pre-LHC data, validated by its successful application to several diffractive processes, is used to predict the total and total-inelastic $pp$ cross sections at the LHC. The prediction for the total-inelastic cross section at $\sqrt s=7$~TeV is compared with recent results from ATLAS and CMS.}

\section{Introduction}
The Froissart bound for the total $pp$ cross section, 
$\sigma_t^{s\rightarrow \infty}<C\cdot (\ln \frac{s}{s_o})^2$ 
(where $s$ is the $pp$ collision energy squared, $C$ is a constant and $s_o$ a scale parameter), which was published fifty years ago~\cite{ref:Froissart} created a keen  interest among the physics community as well as a controversy, which continue to this date. Among the reasons for the continuing interest, as an example, is the possibility of using the optical theorem that relates $\sigma_t$ to the imaginary part of the forward elastic scattering amplitude, ${\rm Im} f_{el}|_{t=0}$, where $t$ is the 4-momentum transfer squared, and dispersion relations that relate the imaginary to the real part, ${\rm Re} f_{el}|_{t=0}$, coupled with a measurement of $\rho={\rm Re} f_{el}|_{t=0}/{\rm Im} f_{el}|_{t=0}$, to look for violations as signs for new physics~\cite{ref:Khuri}. On the other hand, the controversy stems from the coefficient $C$, which was set to $C=\pi/m^2_\pi$ in 1966~\cite{ref:Martin1966}, using $s_o=1$~(GeV$/c)^2$, and updated to $C=\frac{1}{2}\pi/m^2_\pi$ in 2009~\cite{ref:Martin2009}. With such large values of $C$, the bound is more than 100 times higher than the $\sigma_t$ measured at Tevatron energies and in cosmic ray experiments at higher energies, rendering the form of $\sigma_t(s)$ and extrapolations to LHC subject to phenomenological modeling feeding the controversy. 

Measuring cross sections at the LHC involve Monte Carlo (MC) simulations to establish triggers, perform underlying event (UE) corrections and calculate detector acceptances. In anticipation of LHC measurements, MC tuning was intensified and is presently continuing with no ``light at the end of the tunnel'' seen in the search for a MC model that could reliably accommodate all diffractive processes. The present paper is based on a QCD inspired model (RENORM) that addresses all diffractive processes and final states. 

RENORM predictions have been previously presented in Ref.~\cite{EDS-2009} (June 2009) and updated in Ref.~\cite{Stefano} (May 2010). The 2010 paper~\cite{Stefano} represents  a concise summary of the talk delivered at the present conference, and the reader is referred to that paper for details and for the proposed MC strategy for the LHC. In the present paper, we will focus on an update of our model to include a prediction of the total-inelastic cross section, $\sigma_{inel}$.

This update was motivated by the preliminary results for  $\sigma_{inel}$ at $\sqrt s=7$~TeV at the LHC released by ATLAS in February 2011~\cite{ref:sigmainel_ATLASnote}. As the measurement of $\sigma_{inel}$ involves an extrapolation from a ``visible'' to the total-inelastic cross section using MC simulations, and due to the interest in using $\sigma_{inel}$ to measure/monitor the machine luminosity, the simulation of diffractive processes has gained popularity among particle and machine physicists alike. This interest was spread out into the entire particle physics community due to the need to understand the contributions of the diffractive processes to the UE, which affects all measurements at the LHC. 

Below, in Sec.~\ref{Sec:total_elastic_inelastic}, we discuss our predictions for $\sigma_t$, $\sigma_{el}$ and $\sigma_{inel}$ for various values of $\sqrt s$ at the LHC, and in Sec.~\ref{sec:conclusion} we conclude. 

\section{The total, elastic and total-inelastic cross sections\label{Sec:total_elastic_inelastic}}

The elastic, total and single-diffractive (SD) $pp$ cross sections are usually described by Regge theory (see, e.g., Ref.~\cite{myPhysRep}). At high energies, they are  dominated by Pomeron ($\pom$) exchange, and for a Pomeron intercept $\alpha(0)=1+\epsilon$ the $s$-dependence has a power law behavior,
\begin{equation}
\label{eq:xsections}
\left({d\sigma_{el}}/{dt}\right)_{t=0}\sim \left({s}/{s_o}\right)^{2{ \epsilon}},\;\;
\sigma_t=\beta^2_{\pom pp}(0) \cdot\left({s}/{s_o}\right)^{ \epsilon},\;\;\mbox{and}\;
\sigma_{sd}\sim \left({s'}/{s_o}\right)^{2{ \epsilon}},
\end{equation}
where $s'=M^2=s\xi$, $M$ is the mass of the diffractive system and $\xi$ is the forward momentum loss of the proton.
As $s$ increases, this would lead to unitarity violations when the elastic and/or single SD cross section would exceed $\sigma_t$. In the case of SD, CDF measurements   at $\sqrt s=540$~GeV [1800~GeV] showed that a violation of unitarity is avoided by a suppression of $\sigma_{sd}(s)$  by a factor of ${\cal{O}}(5)$ [factor of ${\cal{O}}(10)$] relative to Regge expectations (see Ref.~\cite{KG-95}).

Theoretical models predicting cross sections at the LHC must satisfy necessary unitarity constraints. Unitarization procedures employed by different authors differ in concept and in the number of parameters used that need to be tuned to available accelerator and cosmic ray data. While a rise of the total cross section from Tevatron to LHC is generally obtained, the predictions for LHC are spread out over a wide range. For example, in Ref.~\cite{EDS-2009}, authors predict a $\sigma_t$ at $\sqrt s=14$~TeV ranging from 90 to 250 mb. 
The inherently unitarized RENORM model is based on a saturated Froissart bound and is only subject to uncertainties propagated from the uncertainties in two experimentally determined parameters: a scale parameter $s_o$, and a saturation $s$-value $s_F$ above which the Froissart bound is reached.   

The model has been justified in a recent paper~\cite{myPRD}, where it was introduced as a special phenomenological interpretation of the parton model for the Pomeron in QCD discussed in Ref.~\cite{Levin}. This model is based on wee-parton cascades and yields formulae similar in form to those of Regge theory. Interpreting the term which is equivalent to the Pomeron flux in this model as a gap formation probability, naturally leads to the concept of renormalization as a procedure that eliminates overlapping rapidity gaps in an event, which otherwise would be counted as additional events. The overlapping rapidity gaps are precisely those  responsible for the $s^{2\epsilon}$ factor in SD and elastic scattering in the Regge picture, and would lead to a unitarity violation in the absence of any unitarization.


In Ref.~\cite{Stefano}, the saturated Froissart bound above $s=s_F$ leads to a cross section of the form: 
\begin{equation}
\sigma_t(s>s_F)=\sigma_t(s_F)+(\pi/s_o)\cdot \ln^2(s/s_F).
\label{eq:Froissart}
\end{equation}  
The parameter $s_F$ is determined from the position of a {\em knee} observed in the energy dependence of $\sigma_{sd}$ at $\sqrt s=\sqrt s_{\rm knee}$ (see Fig.~1 in Ref.~\cite{Stefano}). The knee is attributed to a saturation in multiple wee-parton exchanges, manifesting as the scaling parameter $s_o$ of the sub-energy-squared of the diffractive system, $s'\equiv M^2$ (see Eq.~\ref{eq:xsections}), which identifies $s_o$ as a mass-squared, $s_o\equiv M_o^2$. Thus, $M_o$ is reasonably interpreted as the mass of a saturated partonic glueball-like exchange, whimsically named {\em superball} in Ref.~\cite{EDS-2009}. Inserting $s_o$ into Eq.~(\ref{eq:Froissart}) in place of $m^2_\pi$ yields an analytic expression for the total cross section for $s>s_F$.

Predicting the total cross section at the LHC using Eq.~(\ref{eq:Froissart}) 
requires knowledge of $\sigma_t(s_F)$. 
The cross section at $\sqrt{s_F}=22$~GeV, however, has substantial Reggeon-exchange contributions, and also contributions from the interference between the nuclear and Coulomb amplitudes. A complete description must take into consideration all these contributions, using Regge or parton-model amplitudes to describe Reggeon exchanges, and dispersion relations to obtain the real part of the amplitude from measured total cross sections up to Tevatron energies. In the RENORM model, we follow a  strategy that bypasses all these hurdles. For completeness, we outline below all the steps in the cross section evaluation process:

\begin{enumerate}[(i)]
\addtolength{\itemsep}{-0.65em}       
\item Use the Froissart formula as a {\em saturated}\, bound;
\item Eq.~(\ref{eq:Froissart}) should then describe the cross section above the {\em knee} in $\sigma_{sd}$ vs $\sqrt s$, which occurs  at $\sqrt s_F=22$~GeV, and therefore should be valid at the Tevatron at $\sqrt s=1800$~GeV;
\item replace $m_\pi^2$ by $m_{superball}^2=s_o/(\hbar c)^2\approx (3.7\pm 1.5)/ 0.389$ mb$^{-1}$ in the coefficient $C=\pi/m_\pi^2$;
\item note that Reggeon-exchange contributions at $\sqrt s=1800$~GeV are negligible (see Ref.~\cite{CMG-96});
\item obtain the total cross section at the LHC as:
\begin{equation}
\sigma_t^{\mbox{\rm\sc lhc}}=\sigma_t^{\mbox{\rm\sc cdf}}+
{\frac{\pi}{s_o}}
\left[\left(\ln\frac{s^{\rm LHC}}{s_F}\right)^2-\left(\ln\frac{s^{\rm CDF}}{s_F}\right)^2\right].
\end{equation}
\end{enumerate}
Using the CDF $\sigma_t^{\rm CDF}=80.03\pm2.24$~mb at $\sqrt s=1.8$~TeV, this formula predicts the cross sections shown in Table~\ref{tab:x-sections}. The values for $\sigma_{el}$ and $\sigma_{inel}$ are also shown, obtained using the ratios of $R_{el/t}\equiv\sigma_{el}/\sigma_t$ of the global fit of Ref.~\cite{CMG-96}.    
\begin{table}[h]
\caption{Predicted $\sigma_t$, $\sigma_{el}$ and $\sigma_{inel}$ {\em pp} cross sections [mb] at LHC; uncertainties are dominated by that in $s_o$.}
\begin{center}
\begin{tabular}{rlll}\hline\hline
$\sqrt s$ & $\sigma_t$    & $\sigma_{el}$    & $\sigma_{inel}$ \\
\hline
7 TeV     & $98\pm 8$     & $27\pm 2$        & $71\pm 6$ \\
8 TeV     & $100\pm 8$    & $28\pm 2$        & $72\pm 6$ \\
14 TeV    & $109\pm12$    & $32\pm 4$        & $76\pm 8$ \\
\hline\hline
\end{tabular}
\label{tab:x-sections}
\end{center}
\end{table}
 The result for $\sigma_t$ at $\sqrt s=14$~TeV falls within the range of cross sections predicted by the various authors in Ref.~\cite{EDS-2009}, 
and is in good agreement with the value of $114\pm 5\;\rm{mb}$ of the global fit of Ref.~\cite{CMG-96}, where the uncertainty was propagated from the 
$\pm \delta\epsilon$ value reported in the paper using the correlation between $\sigma_t$ and $\epsilon$ through $\sigma_t\sim s^\epsilon$. 

The February 2011 (pre-Moriond) ATLAS result for $\sqrt s=$ 7~TeV was~\cite{ref:sigmainel_ATLASnote}:
\begin{equation}
\sigma_{inel}(\xi>10^{-5})=57.2\pm0.1({\rm stat.})\pm0.4({\rm syst.})\pm6.3({\rm Lumi})\;{\rm mb}
\end{equation}
Based on a PYTHIA (PHOJET) extrapolation, a $\sigma_{inel}=63.3\pm7.0$~mb ($60.1\pm6.6$~mb) was obtained. These results/predictions provided the motivation for updating the RENORM prediction and presenting the result in Moriond-2011.

After Moriond-2011, ATLAS reported the following results from an updated analysis~\cite{ref:sigmainel_ATLASarXiv}:
\begin{eqnarray}
\sigma_{inel}(\xi>10^{-6})=60.33\pm 2.10({\rm exp.})\pm0.4\;{\rm mb}\\
\sigma_{inel}(\xi>m_p^2/s)=69.4\pm2.4({\rm exp.})\pm6.9({\rm extr.})\;{\rm mb}
\end{eqnarray}

Also after Moriond-2011, CMS reported a measurement~\cite{ref:sigmainel_CMS} of $\sigma_{vtx}^{inel}$ based on events with 3 or more particles with $p_T>200$~MeV$/c$ in $|\eta|<2.4$, $\sigma_{vtx}^{inel}=59.7\pm0.1({\rm stat.})\pm 1.1({\rm syst.})\pm2.4({\rm Lumi})$~mb, and using MC models to extrapolate to $\sigma_t^{inel}$ obtained:
\begin{equation}
66.8\leq \sigma_t^{inel}\leq 74.8\;{\rm mb}. 
\end{equation}
Both the ATLAS and CMS results are in good agreement with the RENORM prediction.

\section{Conclusion\label{sec:conclusion}}
The total $pp$ cross section at the LHC is predicted in a phenomenological approach that obeys all unitarity constraints. The approach is based on a saturated Froissart bound above a $pp$ collision energy-squared $s=s_F$, leading to an analytic $\ln^2 (s/s_F)$-dependence, $\sigma_t=(\pi/s_o)\cdot \ln^2(s/s_F)$. The scale parameters $s_F$ and $s_o$ are experimentally determined from pre-LHC SD results. Using the ratio $R_{el/t}\equiv\sigma_{el}/\sigma_t$ from a global fit to cross sections~\cite{CMG-96} to extract $\sigma_{el}$ from $\sigma_t$, a $\sigma^{model}_{inel}=71\pm6$~mb at $\sqrt s=7$~TeV is obtained, which is in agreement with the ATLAS $\sigma_{inel}(\xi>m_p^2/s)=69.4\pm2.4({\rm exp.})\pm6.9({\rm extr.})\;{\rm mb}$ 
and the CMS $66.8\leq \sigma_t^{inel}\leq 74.8\;{\rm mb}$ results.

\section*{Acknowledgments}
Warm thanks to The Rockefeller University and the U.S. Department of Energy Office of Science for financial support, and to my colleagues at Rockefeller, CDF and CMS for many discussions.

\end{document}